%% file: 0_main.tex
\RequirePackage{fix-cm}

\documentclass[twocolumn,epjc3,hidelinks]{svjour3}  

\smartqed

\journalname{Eur. Phys. J. C}

\input{definitions}
\usepackage[switch]{lineno}

\begin{document}


\title{Procurement and Purification of Liquid Argon for the LEGEND-200 Experiment
}

\author{
    Małgorzata Harańczyk\orcidlink{0000-0002-3841-4108} \thanksref{uj,lngs}$^\mathrm{,b}$
   \and
    Patrick Krause\orcidlink{0000-0002-9603-7865} \thanksref{tum}
    \and
    Tomasz Mróz\orcidlink{0000-0002-5304-5531} \thanksref{uj,nm}
    \and
    Laszlo Papp\orcidlink{0000-0002-5221-3548} \thanksref{tum}
    \and
    Krzysztof Pelczar\orcidlink{0000-0001-9504-1750} \thanksref{uj,np}
    \and
    Stefan Schönert\orcidlink{0000-0001-5276-2881} \thanksref{tum}
    \and
    Mario Schwarz\orcidlink{0000-0002-8360-666X} \thanksref{tum}$^\mathrm{, c}$
    \and
    Christoph Vogl\orcidlink{0000-0001-9934-5401} \thanksref{tum}$^\mathrm{, d}$
    \and
    Grzegorz Zuzel\orcidlink{0000-0001-5898-2658} \thanksref{uj}$^\mathrm{, a}$
    \and
     Marco Balata\orcidlink{0000-0001-6745-6983} \thanksref{lngs}
    \and
    Nina Burlac\orcidlink{0000-0002-9877-6266} \thanksref{lngs}
}

\thankstext{e1}{e-mail: \href{mailto:grzegorz.zuzel@uj.edu.pl}{grzegorz.zuzel@uj.edu.pl}}
\thankstext{e2}{e-mail: \href{mailto:malgorzata.haranczyk@uj.edu.pl}{malgorzata.haranczyk@uj.edu.pl}}
\thankstext{e3}{e-mail: \href{mailto:mario.schwarz@tum.de}{mario.schwarz@tum.de}}
\thankstext{e4}{e-mail: \href{mailto:christoph.vogl@tum.de}{christoph.vogl@tum.de}}

\thankstext{nm}{Presently at the Henryk Niewodniczanski Institute of Nuclear Physics, ul. Radzikowskiego 152, 31-342 Krakow, Poland}
\thankstext{np}{Presently at the European Commission, Joint Research Centre (JRC), Geel, 2440, Belgium}

\institute{M. Smoluchowski Institute of Physics, Jagiellonian University, Kraków, 30-348, Poland \label{uj}
\and
Department of Physics, TUM School of Natural Sciences, Technical University of Munich, 85748 Garching, Germany \label{tum}
\and
Istituto Nazionale di Fisica Nucleare, Laboratori Nazionali del Gran Sasso, 67100 Assergi (AQ), Italy \label{lngs}
}

\date{Received: date / Accepted: date}

\maketitle

\begin{abstract}
\legend-200 requires high-purity liquid argon for effective background discrimination. In this paper, we present the design, construction, and performance of a dedicated liquid argon purification system, along with the procurement and purification of liquid argon for filling the \legend{}-200 cryostat to its total capacity of \SI{91}{\tonne}. The purifier is based on copper ca\-ta\-lyst and molecular sieve to remove oxygen and water. Starting with liquid argon of 5.5 quality, featuring an effective scintillation light triplet lifetime $\tau_t$ of about \SI{0.9}{\micro\second}, we achieved a final purity  corresponding to $\tau_t$ = \SI{1.3}{\micro\second}. After complete filling of the \legend{}-200 cryostat, the measured effective triplet lifetime was \SI{1.16}{\micro\second}. The notable reduction is caused by a residual nitrogen impurity introduced by an accidentally spoiled liquid argon delivery. An excessive nitrogen influx was prevented by the \textsc{Legend} Liquid Argon Monitoring Apparatus (LLAMA), which served as one of the three independent purity monitors during the filling campaign.

\vskip 1.0 cm
\end{abstract}

\input{1_introduction}
\input{2_qualities_tests}
\input{3_production_plant_tests}
\input{4_llars_design}
\input{5_L200_filling}
\input{6_conclusions}

\section*{Acknowledgments}

This work was supported by the Polish National Science Centre (UMO-2020/37/B/ST2/03905), the Polish Ministry of Science and Higher Education (2022/WK/10), the BMFTR through the Verbundforschung 05A20WO2, and by the Excellence Cluster ORIGINS, which is funded by the Deutsche Forschungsgemeinschaft (DFG, German Research Foundation) under Germany’s Excellence Strategy – EXC 2094 – 390783311.

The Authors would like to thank Natalia Di Marco, Giuseppe Salamanna, and Diego Tagnani for their help and assistance in operating the purification system during the filling of the \legend-200 cryostat. 

\printbibliography
\end{document}

%% file: 1_introduction.tex
\section{Introduction}
\label{intro}

The \legend{} Collaboration aims for an unambiguous discovery of neutrino-less double beta (0$\nu\beta\beta$) decay of $^{76}$Ge. 
This lepton-number-violating process, if observed, would provide direct proof that neutrinos are Majorana fermions. 
The decay rate will also allow constraining the absolute neutrino mass scale and support theories in which leptons contributed to the observed matter-antimatter asymmetry in the present Universe. 
It would therefore be a critical contribution to modern particle, nuclear, and astrophysics.

The \legend{} experiment applies a staged approach to achieve its scientific objectives. 
Currently, the \legend-200 stage~\cite{legendcollaborationFirstResultsSearch2025} is running in Hall A of the Laboratori Nazionali del Gran Sasso (LNGS) underground laboratory in Italy. 
It paves the way for the next generation \legend-1000 stage~\cite{legend1000-pcdr}, which will be located in Hall C of LNGS.

The experiment uses high-purity germanium detectors enriched in $^{76}$Ge, suspended in liquid argon (LAr). 
Apart from cooling the detectors and passively stopping external radiation, the scintillating LAr serves as an active shield against backgrounds and features a light readout instrumentation to this end. 
This use of LAr is adopted from \gerda~\cite{gerda-final}. 
Along with \textsc{Majorana Demonstrator}~\cite{mjd-technical}, they are predecessors of \legend. 
As in \gerda, the LAr instrumentation is an obligatory building block for achieving a quasi-background-free search, which is required to meet the science goal.

A higher exposure goal necessitates lower background levels: 
\legend-200 aims for a background index of \SI{2e-4}{\ckky}, which is a factor of 2.5 lower than what was achieved in \gerda{} Phase-II. 
This calls for improvements across all background-reduction techniques, including the LAr scintillation detector.
Excellent optical properties of LAr are essential to enhance the LAr anti-coincidence performance, foremost the primary light yield (i.e., the number of photons created at the interaction point per unit energy) and the attenuation length. Both parameters deteriorate due to impurities present in LAr.

\gerda's cryostat was filled with LAr of 5.0 purity\footnote{Though certified as 5.0, \gerda's LAr turned out much cleaner than the allowed limits.} in 2009 \cite{Knopfle:2022fso};
the LAr retained a reasonable purity until the end of data taking in 2019, without any re-purification \cite{Wiesinger:PhD}. 
For \legend-200, we aimed for a higher initial purity by enforcing stronger limits on delivered LAr and performing liquid-phase purification with a dedicated device during filling.
Three independent purity monitors assured high quality throughout the filling process.

This paper introduces two devices for measuring LAr purity in Sec.~\ref{sec:LAr_qualities_tests}, presents tests of the LAr delivery chain to LNGS in Sec.~\ref{sec:LAr_qualities_tests_plant}, and shows the design of the \legend-200 LAr purification system in Sec.~\ref{sec:design}. Finally, we report on the system's performance during filling the \legend-200 cryostat in Sec.~\ref{sec:filling}.

%% file: 2_qualities_tests.tex
\section{Purity measurements of commercially available liquid argon}
\label{sec:LAr_qualities_tests}

\subsection{Impurities in liquid argon}
\label{subsec:impure}

High-quality liquid argon is defined by a low impurity content. Impurities in liquid argon deteriorate its optical properties by quenching the scintillation light emission and absorbing emitted photons. The quenching process reduces the primary light yield and the effective triplet lifetime \cite{warp-oxygen}.

The quality of liquid argon can be assessed by measuring the concentration and composition of impurities or by measuring the liquid's optical properties.
Our strategy in this work encompasses both options.

\subsection{Argon purity monitors}
\label{subsec:LAr_qualities_tests_monitors}
Three separate and complementary devices were utilized to ensure the highest possible purity of liquid argon filled into the \legend-200 cryostat. Two external purity monitors were employed: the mobile scintillation analyzer (SA), which measures the effective argon triplet lifetime parameter in liquid argon samples, and the gas analyzer (GA), which directly measures impurity levels in evaporated gaseous argon. They were used as monitors to establish the \legend-200 LAr delivery chain, as well as during the design and commissioning process of the \legend-200 Liquid Argon Purification System (LLArS). 
The third apparatus, \llama{}~\cite{schwarzPhD}, is installed inside the \legend-200 cryostat to monitor the argon purity during operation of \legend-200. All three monitors were employed during the cooling-down and filling of the \legend-200 cryostat with purified liquid argon. This section presents the qualities of both external purity monitor devices. LLAMA is described in Section \ref{sec:llama} as a part of the cryostat filling operation.

\subsubsection{Mobile Scintillation Analyzer}
\label{subsec:SA}

\begin{figure}[b]
\centering
  \includegraphics[width=1.0\columnwidth]{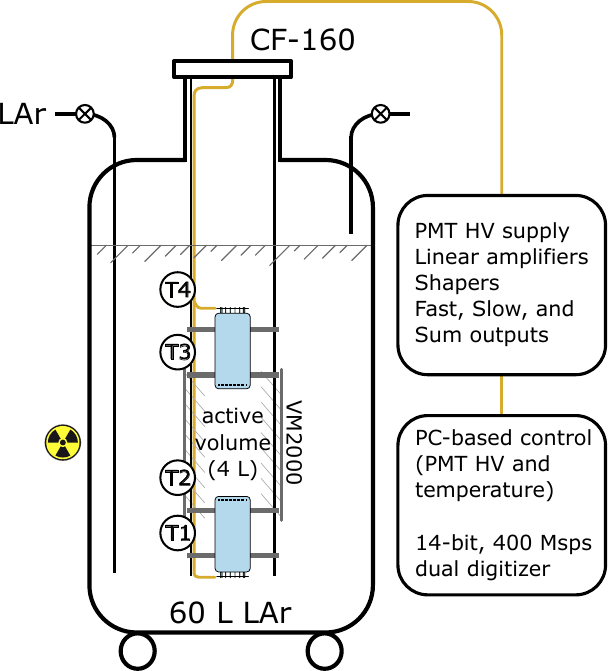}
\caption{The mobile scintillation analyzer. The base of the system is a \SI{60}{\litre} fully metal-sealed dewar with a wide neck closed by a CF-160 flange. On the flange, an assembly of two TPB-coated \SI{2}{\inchQ} PMTs is installed. The PMTs observe approximately \SI{4}{\litre} of LAr, surrounded by a VM2000 multi-layer reflector foil. The fill level is controlled by four Pt-100 temperature sensors. The integrated spectroscopy chain unit conditions the signals and provides high-voltage bias to the PMTs. A two-channel, 14-bit, \qty{400}{Msps} digitizer records PMT signals. Natural radioactivity is sufficiently strong to perform measurements; however, a dedicated external gamma source may be used for additional LAr irradiation.}
\label{fig:PMT_dewar}     
\end{figure}

A schematic diagram of the mobile scintillation analyzer is shown in  Fig.~\ref{fig:PMT_dewar}.
The SA was constructed at the Jagiellonian University in Kraków and is dedicated to measuring relative changes in the effective triplet lifetime of LAr before and after LAr purification. 

\begin{figure*}[t]
\centering
\begin{annotate}{\includegraphics[width=0.98\textwidth]{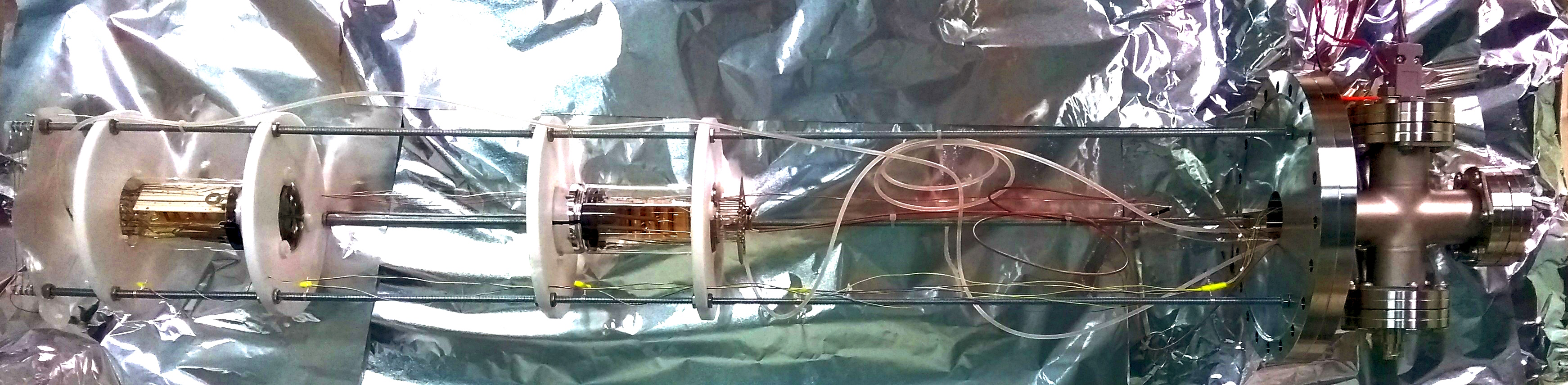}}{1}
    \callout{ 3.3, 2.1}{Signal cables}{ 3.3, 0.0}
    \callout{ 1.5, 2.1}{HV cables}{ 1.5, 0.6}
    \callout{-1.5, 2.1}{Top PMT}{-1.5, 0.5}
    \callout{-6.5, 2.1}{Bottom PMT}{-6.5, 0.5}
    \callout{-4.5, 2.1}{Active volume}{-4.5, 0.9}
    \draw[thick,red,dashed] (-5.8, 0.8) rectangle (-2.6, -1.3);

    \callout{-7.0, -2.1}{T1}{-7.0, -1.2}
    \callout{-5.2, -2.1}{T2}{-5.2, -1.2}
    \callout{-2.2, -2.1}{T3}{-2.2, -1.2}
    \callout{ 0.2, -2.1}{T4}{ 0.2, -1.2}
    
    \callout{ 6.4,  2.1}{Top flange and feedthroughs}{ 6.4,  0.8};
\end{annotate}
\caption{Cryogenic photomultiplier assembly. Two \SI{2}{\inchQ}, TPB-coated photomultipliers are placed about \SI{20}{\centi\meter} apart, facing each other. The active volume is enclosed by VM2000 multi-layer reflector foil. Four Pt-100 temperature sensors (T1--T4) are located above the voltage dividers and photocathodes of the bottom and top PMTs. Temperature readouts indicate the cryogenic liquid level.}
\label{fig:PMT_assembly}     
\end{figure*}

The SA is based on a \SI{60}{\litre} metal-sealed dewar, containing two \SI{2}{\inchQ} photomultiplier tubes (PMTs) equipped with voltage dividers. The PMTs are coated with Tetraphenyl-Butadiene (1,1,4,4-tet\-ra\-phe\-nyl-1,3-bu\-ta\-die\-ne, TPB) since detection of liquid argon scintillation light requires a wavelength shifter (WLS).
The WLS absorbs \SI{128}{\nano\meter} scintillation photons and re-emits visible photons, detectable by the PMTs. Signals from the PMTs are amplified by dedicated fast amplifiers, which can also shape pulses. The amplifier circuit enables recording data from both PMTs independently or acquiring summed pulses. A two-channel, 14-bit, \qty{400}{Msps} digitizer M3i.4142 by Spectrum Instrumentation ~\cite{SpectrumM} records PMT signals, and dedicated software makes it possible to determine the effective triplet lifetime in real time to allow for a fast judgment about the investigated liquid argon quality. Data is also stored on disk for further offline analysis. A detailed picture of the PMT assembly construction is shown in Fig.~\ref{fig:PMT_assembly}.

The effective triplet lifetime fitting method assumes a simplified model of scintillation light emission.  Detailed analysis of delayed TPB re-emission effects~\cite{segreto} was ignored in the analysis, assuming that the dominant TPB re-emission is prompt. To accommodate this simplification, the model introduces an effective intermediate scintillation term covering both argon molecule recombination~\cite{hofmann} and delayed TPB re-emission. As a result of these assumptions, the model describes the argon scintillation time profile using three distinct components:
\begin{equation}
    A(t, \tau_s, \tau_i, \tau_t) = \frac{A_s}{\tau_s}e^{\frac{-t}{\tau_s}} + \frac{A_i}{\tau_i}e^{\frac{-t}{\tau_i}} + \frac{A_t}{\tau_t}e^{\frac{-t}{\tau_t}},
    \label{eq:argon_profile}
\end{equation}
where $\tau_s$ ($\tau_i$, $\tau_t$) is the singlet (intermediate, triplet) decay times, and $A_s$ ($A_i$, $A_t$) are constants related to the intensities of singlet (intermediate, triplet) states. The LAr scintillation time profile is then convolved with the Gaussian-modeled temporal response of the detection system, including light propagation effects. The resulting fit function is the sum of exponentially modified Gaussian functions. 

\begin{figure}[htb]
\centering
  \includegraphics[width=1.0\linewidth]{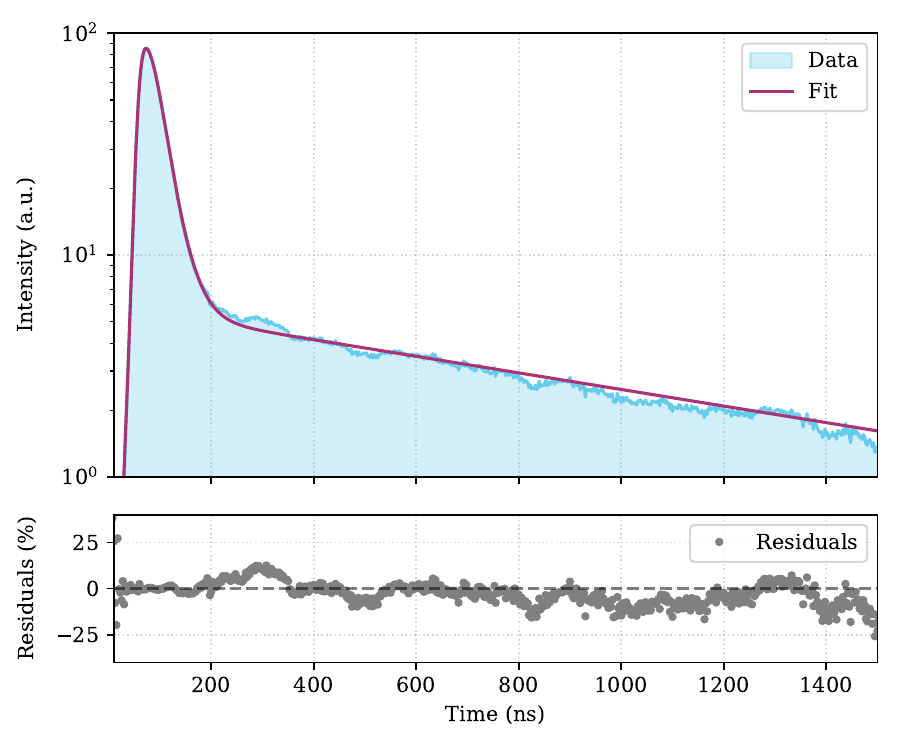}
\caption{Exemplary super-pulse from data acquired with both PMTs summed up. The fit returns an effective triplet lifetime value of $\tau_t = \SI{1.16(3)}{\micro\second}.$}
\label{fig:tau_fit}     
\end{figure}

The model is fitted to an aggregate of many pulses, forming a super-pulse.
The pulses may be collected from individual PMTs or from the sum of signals from both PMTs. An example of a fit to data acquired from the top PMT for liquid argon 5.0 is shown in Fig.~\ref{fig:tau_fit}.

\subsubsection{Gas Analyzer}\label{sec:gas_analyzer}

\begin{figure}[thb]
\centering
  \includegraphics[width=\linewidth]{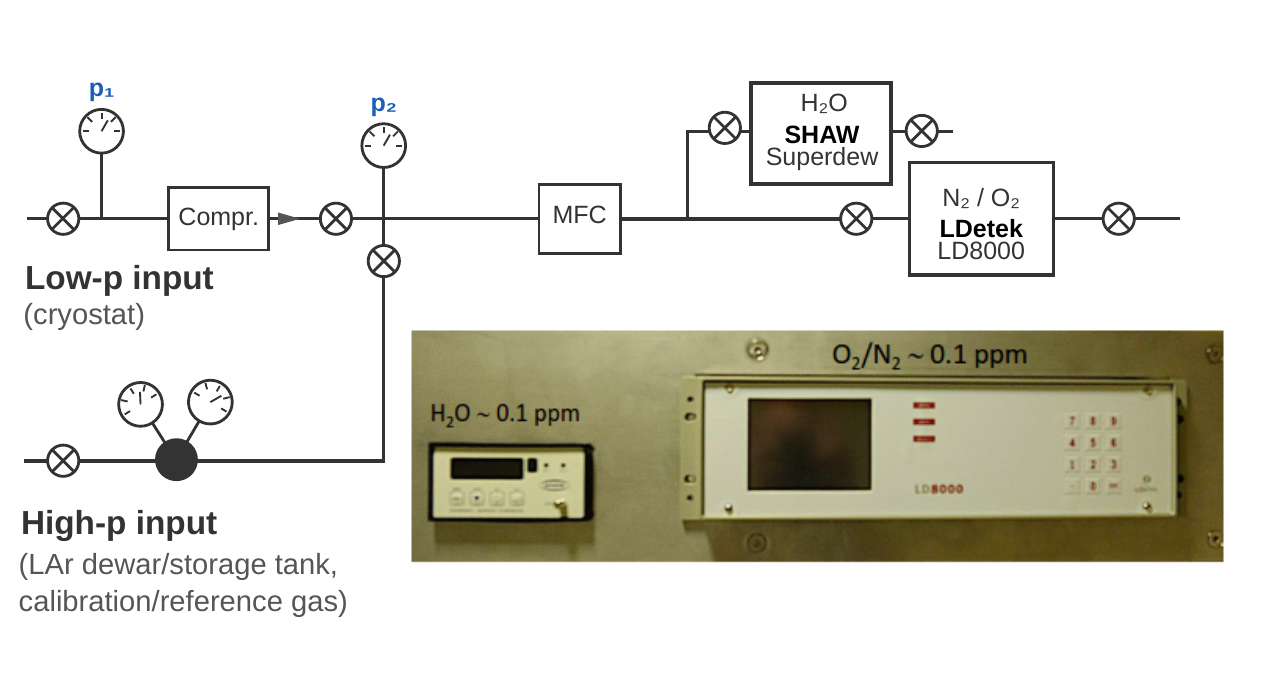}
\caption{Scheme and a picture of a panel with the Gas Analyzer components. The dew point meter and the \nit/\oxy{} analyzers are connected in parallel. If the gas source to be analyzed is not pressurized, an input with a dedicated compressor may be used (low-p input). The high-pressure (high-p) input is equipped with a high-purity pressure reducer.}
\label{fig:gas_scheme}     
\end{figure}

A second device called gas analyzer has been constructed to enable continuous monitoring of the concentrations of nitrogen, oxygen and water in argon gas with a sensitivity of around \qty{0.1}{ppm}\footnote{$\qty{1}{ppm} = \qty{1}{ppm(v)} = \qty{1e-6}{\cubic \cm \per \cubic \cm}$}. 
The system is based on the following commercial components: the LDetek LD8000 multigas trace impurity analyzer and the SHAW Superdew 3 Hygrometer with the Silver sensor. 
The Silver sensor can measure the dew point of process gases and compressed air. 
Its specific range is \qty{-100}{\degreeCelsius} to \qty{-20}{\degreeCelsius} dew point, corresponding to a volume fraction of \qty{0}{ppm} to \qty{1000}{ppm}. 
A schematic diagram of the GA is shown in Fig.~\ref{fig:gas_scheme}. 
The LDetek LD8000 measures nitrogen and oxygen in gaseous argon. Nitrogen is detected by a plasma emission detector, and oxygen by an electrochemical cell. Both sensors feature individual electronic flow controllers with minimal dead volumes, ensuring a fast response time. 
The analyzer has a built-in getter that produces a reference zero-gas using Ar 6.0 as input to determine the background. 
Calibration of the instrument was performed with a certified calibration gas mixture based on Ar 6.0 and containing a nominal amount of \qty{10}{ppm} nitrogen and \qty{10}{ppm} oxygen. 
The device is designed for argon monitoring with typical flows of \qty{150}{\cubic \cm \per \minute} and requires approximately \qty{12}{hours} of purging time before reaching nominal sensitivity.
The delay is primarily due to the oxygen sensor being saturated with environmental air. 
The minimum input pressure is about \qty{0.2}{barg} and a typical operational value is about \qty{0.7}{barg}. 
If the input pressure is too low, a dedicated, fully metal-sealed compressor can be used.
For a high-pressure source, the high-p input can be used, as it is equipped with a high-purity pressure reducer. All the connecting tubes are stainless steel, and all the connectors and joints are metal-sealed. 
Properly controlling the gas flow is critical for successful calibration.
Therefore, the device is equipped with a mass flow controller (MFC).

The dew point meter and the LD8000 are connected in parallel, allowing for individual adjustments of gas flow for each detector.

According to the manufacturer, the accuracy of the LD8000 is around \qty{0.1}{ppm}.
The dew point meter features an uncertainty of \qty{2}{\celsius} which translates to an uncertainty on the moisture content $c$ of $\Delta c \approx 0.29 c + \qty{0.02}{ppm}$ for $c$ below \qty{5}{ppm}.

\subsection{Purity of commercial liquid argon}
To investigate the scintillation properties of commercially available argon, twelve samples were collected from different suppliers and delivered in various transport vessels with volumes not exceeding \SI{400}{\litre}. All samples were certified Ar 5.0 or 6.0 qualities. The samples originated from the INFN-LNGS laboratory delivery chains, from the Jagiellonian University in Kraków, Poland (IFUJ) source, and from a dedicated delivery ordered by the \legend{} Collaboration. 

The effective triplet lifetime $\tau_t$ was measured for all samples using the scintillation ana\-ly\-zer described in Sec.~\ref{subsec:SA}. The results are presented in Tab.~\ref{tab:triplet_tests}. 
The uncertainty of $\tau_t$ is about \qty{5}{\percent} of the quoted values. The measured values for Ar 5.0 are within the range of \qtyrange{600}{800}{\ns}.
Surprisingly low values were obtained for Ar 6.0; there, $\tau_t$ is comparable to the measurements performed with Ar 5.0.
The short, effective triplet lifetimes observed in Ar 6.0 samples could, however, be explained by the fact that the gas was delivered in transport vessels equipped with pressure build-up systems.
These are not completely tight and could have allowed air to contaminate the contained argon.

The high variance in the measured effective triplet lifetime for the same gas quality was a key motivation for a more detailed investigation of the production and delivery chain described in Sec.~\ref{sec:LAr_qualities_tests_plant}.  

\begin{table}[htbp]
\centering
\caption{Effective triplet lifetimes $\tau_t$ of various commercial liquid argon sources measured with the scintillation analyzer. The last column specifies the sample's origin (delivery chain) and batch number. For some deliveries, the measurements were repeated on different sub-samples.}
\begin{tabular}{ c c c l}
\toprule
Date  & Purity level & $\tau_t$ (\si{\nano\second}) & Delivery \& batch \\ \midrule
09/03/2019  & 5.0 & 815 & IFUJ-1 \\
01/05/2019  & 5.0 & 716 & IFUJ-2 \\
31/05/2019  & 5.0 & 957 & IFUJ-3 \\
01/06/2019  & 5.0 & 925 & IFUJ-3 \\
\hline
19/06/2019  & 6.0 & 619 & LNGS-1 \\
20/06/2019  & 6.0 & 705 & LNGS-1 \\
29/08/2019  & 6.0 & 684 & LNGS-2 \\
\hline
19/09/2019  & 5.0 & 953 & IFUJ-4 \\
\hline
16/10/2019  & 5.0 & 662 & LNGS-3, \legend \\
17/10/2019  & 5.0 & 663 & LNGS-3, \legend \\
18/10/2019  & 5.0 & 658 & LNGS-3, \legend \\
\hline
30/01/2020  & 5.0 & 1125 & IFUJ-5 \\
\bottomrule
\end{tabular}
\label{tab:triplet_tests}
\end{table}

%% file: 3_production_plant_tests.tex
\section{Full delivery chain tests}
\label{sec:LAr_qualities_tests_plant}

\subsection{Tests at the Linde liquid argon production plant in Trieste}

To enhance our understanding of issues related to the production and transportation of LAr, we have visited the Linde LAr production plant in Trieste, Italy. The plant was originally designed to produce LAr of 5.0 quality and later upgraded to deliver 5.5 quality gas. Sometimes, after longer operation runs, the gas can reach 6.0 purity. Produced LAr is pumped into a storage tank with a capacity of \qty{500}{\cubic \m} without the possibility of bypassing it. The production rate may be adjusted between \qty{5.6}{\cubic \m \per \day} and \qty{10.6}{\cubic \meter \per \day}. 
Tankers with typical capacity of \qty{21}{\cubic \m} (\qty{29}{\tonne}) are filled from the storage tank. Linde personnel can monitor the purity of gas in the storage tank, in the tankers (e.g., after filling), or in other external containers equipped with a dedicated connector. A system based on gas chromatography devices is used for this purpose. 
The procedure assumes that a measurement is performed only after at least 20 minutes of purging the line connecting the gas source with the instrument (argon gas flows from the vessel to the gas chromatograph). 
The detection limit of the system is \qty{0.5}{ppm} for nitrogen and oxygen, and \qty{0.9}{ppm} for water. 
Information on uncertainties is not available. 
The measurements at the LAr production site are performed by Linde personnel using their instruments.
 
While visiting Linde, there were only \qty{37}{\cubic \m} of LAr in the storage tank. The gas was produced shortly after a power shutdown at the plant; thus, its quality was expected to be worse than usual. The measured concentrations of nitrogen, oxygen and water were $[\text{N}_2] \leq \qty{0.5}{ppm}$, $[\text{O}_2] = \qty{1.4}{ppm}$, and $[\text{H}_2\text{O}] \leq \qty{0.9}{ppm}$. This corresponds to a purity level between 5.0 and 5.5, as shown in Tab.~\ref{tab:Ar_purity_spec}.

\begin{table}[bhtp]
\centering
\caption{Selected technical Ar purity specifications provided by Linde. The maximum permissible concentrations of the most relevant elements and compounds are given for the respective purity levels.}
\begin{tabular}{lrrrrr}
\toprule
             &       & \multicolumn{4}{c}{Purity levels} \\ \cmidrule{3-6}
  Substance  & Unit  &     4.8   &    5.0  &      5.5  &      6.0  \\ \midrule
  $[\text{Ar}]$ & \unit{\percent}             &  99.998   & 99.999  &  99.9995  &  99.9999  \\ 
  $[\text{N}_2]$ & ppm           &  $\leq 10$  & $\leq 5$    & $\leq 1.5$  &  $\leq 0.5$ \\ 
  $[\text{O}_2]$ & ppm           &  $\leq 3$   & $\leq 2$    & $\leq 1.0$  &  $\leq 0.5$ \\ 
  $[\text{H}_2\text{O}]$ & ppm   &  $\leq 5$   & $\leq 3$    & $\leq 1.0$  &  $\leq 0.5$ \\ 
  $[\text{C}_n\text{H}_m]$ & ppm &  $\leq 0.5$ & $\leq 0.2$  & $\leq 0.1$  &  $\leq 0.1$ \\ 
   \bottomrule
\end{tabular}
\label{tab:Ar_purity_spec}
\end{table}

It is expected that the filling procedure of transport tanks (of any kind) is the most critical step for possible re-contamination of the gas~\cite{bxnitrogen}. 
Therefore, we wanted to investigate how Linde's standard procedures could affect gas purity. 
We started our tests with the refill (performed by Linde personnel) of an empty Linde tanker, which had previously been used to transport LAr 6.0.  Upon arrival, there was no liquid argon in the tanker.
The tanker was pressurized with argon. Analyzing this gas, Linde found only upper limits for the contaminants: $[\text{N}_2] \leq \qty{0.5}{ppm}$, $[\text{O}_2] \leq \qty{0.5}{ppm}$, and $[\text{H}_2\text{O}] \leq \qty{0.9}{ppm}$. 
Next, the truck was connected to the filling station and the connecting flexible line was purged  for \qty{20}{\minute} with gas from the tanker, i.e., in the reverse direction. 
After about \qty{5}{\tonne} (\qty{3.6}{\cubic \m}) of LAr was transferred into the tanker an analysis of the gas in the tanker was performed, yielding $[\text{N}_2] \leq \qty{0.5}{ppm}$, $[\text{O}_2] = \qty{1.4}{ppm}$, and $[\text{H}_2\text{O}] \leq \qty{0.9}{ppm}$. 
The result was consistent with that obtained for the gas in the storage tank. Therefore, one can conclude that the applied filling procedure does not introduce contamination into the gas at a purity level of 5.0 to 5.5. It has also been confirmed that the purity of LAr in a tanker can be verified and certified.
An overview of the measurements is provided in Tab.~\ref{tab:linde_trieste_tests}. 
\begin{table}[bhtp]
\centering
\caption{Tanker filling test performed at the Linde production plant in Trieste. The measurements demonstrate that a LAr tanker can be filled in a clean way, without contaminating the product.}
\begin{tabular}{cccl}
\toprule
 \multicolumn{3}{c}{Contamination level (ppm)} & {} \\ \cmidrule{1-3}
  \nit   & \oxy   & \water   &  Remarks \\ \midrule
  $\leq$ 0.5  &  1.4   & $\leq$ 0.9   & LAr in storage tank  \\
  $\leq$ 0.5  &  $\leq$0.5   & $\leq$ 0.9   & Ar 6.0 gas in tanker  \\
  $\leq$ 0.5  &  1.4   & $\leq$ 0.9   & LAr in tanker  \\ \bottomrule
\end{tabular}
\label{tab:linde_trieste_tests}
\end{table}

\begin{table}[b]
\centering
\caption{Transport dewar fill test performed at the Linde production plant in Trieste. The first four measurements are related to the boil-off gas sampled from the \qty{180}{\liter} vessel right after it was filled with LAr from the tanker. Subsequent measurements were taken after additional flushing of the Linde instrument. To verify the apparatus, gas from the tanker was sampled. In the final block of the table, the boil-off gas and argon evaporated from the liquid from the \qty{180}{\liter} dewar were re-measured.}
\begin{tabular}{cccl}
\toprule
 \multicolumn{3}{c}{Contamination level (ppm)} & {} \\ \cmidrule{1-3}
  \nit   & \oxy   & \water   &  Remarks \\ \midrule
  3.6  &  2.1   & $\leq$ 0.9   & Boil-off gas \\
  4.3  &  2.3   & $\leq$ 0.9   & Instrument flushed for 10 min \\
  7.8  &  3.1   & $\leq$ 0.9   & +10 min flushing \\
  9.0  &  3.3   & $\leq$ 0.9   & +15 min flushing \\
  \hline
  $\leq$ 0.5  &  1.4   & $\leq$ 0.9   & Ar gas from tanker  \\
  \hline
  7.8  &  3.0   & $\leq$ 0.9   &   Boil-off gas \\ 
  1.7  &  1.8   & $\leq$ 0.9   &  Evaporated LAr \\ 
  \bottomrule
\end{tabular}
\label{tab:linde_trieste_tests_dewar}
\end{table}

In the next step, we tested a procedure of filling a \qty{180}{\liter} transport dewar (Taylor-Wharton, model XL-186/AC-180, provided by Linde) from the tanker. 
After connecting, the dewar was thoroughly flushed with argon gas for about 2 hours before it was filled with liquid. An analysis of the gas sampled from the dewar's gas phase (boil-off) showed significantly higher concentrations of nitrogen and oxygen. Additional purging of the measurement device did not improve the situation. Moreover, the concentration of nitrogen and oxygen was constantly increasing, as shown in the first block of Tab.~\ref{tab:linde_trieste_tests_dewar}. 

A cross-check of the instruments and analysis procedure was performed by reconnecting the instrument to the tanker for re-analysis. The result was consistent with the initial measurement of the tanker's gas quality shown in Tab.~\ref{tab:linde_trieste_tests}. 
Final gas-quality measurements were taken from the gas-phase (boil-off) and liquid-phase ports of the transport dewar.  A significantly lower level of contamination was found in the liquid phase compared to the boil-off gas. This may indicate an air leak into the dewar, possibly from the pressure build-up system. We therefore decided against using small transport containers with a pressure build-up system for \legend.

\subsection{Delivery procedure and tests at LNGS}
\label{subsec:LAr_qualities_tests_definition}

Together with Linde, we defined a special LAr quality, referred to as ``L-200 quality'', to be delivered to the \legend{} experimental site at LNGS. The concentrations of oxygen and water must be below \qty{1}{ppm} and the concentration of nitrogen below \qty{1.4}{ppm}. This is the best quality the Linde Trieste plant can provide during standard operations. According to the specification in Tab.~\ref{tab:Ar_purity_spec}, this corresponds to argon class 5.5\footnote{The company claims to be able to tune the plant and produce argon 6.0, if requested.}. Each tanker is required to carry a certificate of purity of the delivered gas.

We also defined a filling procedure for the \legend-200 LAr storage tank, which was followed by Linde personnel. It was based on the procedure applied in the \borexino{} experiment~\cite{bxnitrogen}. It assumes purging of the filling hose with warm\footnote{The purging speed was adjusted to prevent freezing of remaining traces of water.} argon gas for about \qty{30}{\min} from both sides: first from the tanker towards the storage tank, and later from the storage tank to the tanker, releasing the gas through dedicated vent ports. 

The full delivery chain for L-200-quality LAr has been tested. A full tanker loaded with certified LAr arrived at the underground laboratory, and filled the \legend-200 storage tank (\qty{8}{\tonne} capacity) with \qty{2}{\tonne} of LAr according to the established procedure. 
The purity measurements at consecutive steps of the delivery chain are summarized in Tab.~\ref{tab:linde_delivery_test}. 
During the filling of the storage tank, argon purity was continuously monitored in-line using the gas analyzer described in Sec.~\ref{sec:gas_analyzer}.
The results are presented in the second row of Tab.~\ref{tab:linde_delivery_test}. A slightly elevated nitrogen concentration compared to the certificate has been found (\qty{0.8}{ppm} vs.\ the upper limit of \qty{0.5}{ppm}). Also, due to the short purging time of less than an hour, the oxygen sensor of the gas analyzer did not reach its nominal sensitivity and could only report upper limits for this measurement (\qty{0.8}{ppm}). From the test, it can be concluded that the established procedures for the \legend-200 argon delivery maintain high purity and fulfill the experiment's needs.

After filling the storage tank, the purity of the liquid sampled from it was investigated using the scintillation analyzer presented in Sec.~\ref{subsec:SA}.
An effective triplet lifetime of \qty{0.7}{\micro \s} was found. The measured effective triplet lifetime is short and unsatisfactory, demonstrating the need for LAr purification for \legend-200.
Argon sampled from both the storage tank's gas and liquid phases was also tested with the gas analyzer. The elevated nitrogen concentration initially detected during storage tank filling was confirmed. Meanwhile, the oxygen sensor seems to have settled and revealed a significant decrease compared to the certificate. The measurements by the gas analyzer are shown in rows three and four of Tab.~\ref{tab:linde_delivery_test}.

\begin{table}[htbp]
\centering
\caption{Measurements of LAr purity conducted with the gas analyzer. We observed the filling of the \legend-200 LAr storage tank (ST), measured the impurity concentration in the tank after filling was completed, and investigated LAr in the scintillation analyzer (SA) vessel. In step 2, the oxygen sensor was not yet sufficiently purged, yielding only upper limits on the oxygen contamination.}
\begin{tabular}{ccccl}
\toprule
& \multicolumn{3}{c}{Contamination level (ppm)} & {} \\ \cmidrule{2-4}
Step & \nit{} & \oxy{} & \water{} &  Remarks \\ \midrule
1 & $\leq$ 0.5  &  0.9   &  $\leq$ 0.9  & Tanker certificate \\ 
2 & 0.8 & $\leq$ 0.8 & 0.6 & in-line ST filling \\ 
3 & 0.9  &  0.2   &  0.5   & ST liquid \\

4 & 0.6  &  0.2   &  0.3  & ST gas \\

5 & 0.9  &  0.2   &  0.4   & SA liquid \\

6 & 1.2  &  0.2   &  0.2  & SA gas \\
\bottomrule
\end{tabular} 
\label{tab:linde_delivery_test}
\end{table}

Finally, we measured the impurity concentrations of the gas and liquid in the scintillation analyzer (rows five and six in Tab.~\ref{tab:linde_delivery_test}). The liquid in the scintillation analyzer exhibited impurity levels comparable to those in the storage tank, whereas the gas phase contained an excess of nitrogen, suggesting degassing from warm parts. Comparable impurity levels in the liquid from the storage tank and the scintillation analyzer indicate that the SA filling procedure does not affect gas quality.

Clearly, the established delivery chain, refilling procedures, and equipment are capable of delivering and verifying the \legend-200 LAr quality. The developed analyzers are sufficiently sensitive to perform the desired analyses and quality monitoring during the \legend{} cryostat filling operation, given sufficient purging time for the gas analyzer.

%% file: 4_llars_design.tex
\section{The LEGEND-200 Liquid Argon Purification System}
\label{sec:design}

\subsection{Working principle of liquid argon purification}

Commercial argon is mainly contaminated with the principal constituents of air, i.e.\ nitrogen, oxygen, and water.
Oxygen and water can be effectively removed from liquid argon by chemisorption and physisorption on special materials~\cite{purifiers}. Nitrogen is more difficult to remove, but also not as critical for LAr scintillation properties as water or oxygen, since a few ppm of nitrogen can still be tolerated. 
Recent investigations, which have surfaced after the \legend-200 LAr cryostat was filled, have shown that Li-FAU zeolite features excellent nitrogen capturing capabilities~\cite{cardosoInnovativeProposalCapturing2024,bianchinetoExploringN2Capturing2025}. 

Physisorption is a physical process in which molecules bind to porous surfaces via van der Waals forces, without altering their chemical composition. This effect is used to remove water using silica gel or molecular sieves of an appropriate pore size. Chemisorption, on the other hand, involves a chemical reaction between the contaminant and the filter agent.
The chemisorption mechanism is used to capture oxygen on copper via an exothermic reaction,
\begin{equation}\label{eq:copper-oxidation}
    \ce{2Cu + O2 -> 2CuO + heat}.
\end{equation}
This reaction can be reversed, providing the possibility of oxygen trap regeneration by reducing  copper(II) oxide with hydrogen to elemental copper,
\begin{equation}\label{eq:cu-regeneration}
    \ce{CuO + H2 -> Cu + H2O + heat}.
\end{equation}
The reaction in Eq.~\ref{eq:cu-regeneration} is also exothermic, but is usually performed at an elevated temperature to increase the reaction rate.

\subsection{Initial purification tests at Jagiellonian University in Krak\'{o}w}
\label{sec:design_adsorbers}
\begin{figure}[htb]
\centering
  \includegraphics[width=1.0\columnwidth]{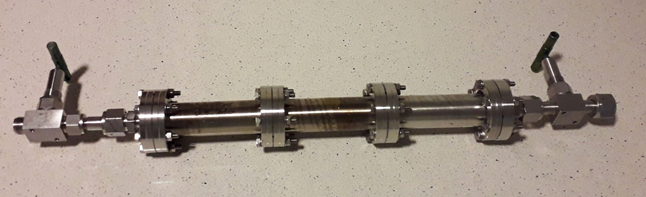}
\caption{Prototype oxygen trap made from 3 CF-40 nipples and containing about \qty{300}{\g} of Puristar R3-11G copper catalyst.}
\label{fig:prototype_trap}     
\end{figure}

For initial proof-of-concept tests performed in Krak\'{o}w, LAr was purified from oxygen only, as it is considered to have the greatest impact on the scintillation properties of LAr. We used the Puristar R3-11G copper catalyst by BASF, which has a specific oxygen capacity of {\num{4.72 (67)}~{norm.}~L/kg} \cite{purifiers}. The catalyst comes in the form of $3 \times \qty{5}{\mm}$ pellets with a target CuO content of \qty{45}{\percent} by weight. The rest are Mg-silicates and promoters. The catalyst was placed in a trap made from 3 CF-40 nipples connected together and closed on both sides with metal-sealed 1/2-inch Swagelok valves, as shown in Fig.~\ref{fig:prototype_trap}. The total mass of the adsorber, approximately \qty{300}{\gram} (\qty{360}{\cubic \cm}), was secured in place by glass wool frits. The Puristar catalyst was activated (regenerated after saturation with atmospheric oxygen) by flushing it with Arcal-15 mixture (Ar + \qty{5}{\percent} hydrogen) at about \qty{250}{\celsius} until around 900 trap volumes have been exchanged and the water content in the exhaust gas dropped to a constant level of a few ppm (monitored). The regeneration procedure is described in detail in Sec.~\ref {subsec:regeneration}. 

Approximately \qty{60}{\liter} of LAr 5.0 were purified using this prototype setup: LAr was pushed from a \qty{400}{\liter} transport tank through the trap into the scintillation analyzer. The effective triplet lifetime measured for the purified argon was \qty{1.3}{\micro \s}. The same LAr batch yielded $\tau_t = \qty{0.8}{\micro \s}$ before purification. This test demonstrated that the 5.0 grade LAr can be effectively purified to achieve the high purity required to scintillate with the nominal triplet lifetime.

\subsection{Design of the LEGEND-200 purification system}
\label{subsec:design_system}

\begin{figure*}[htbp]
\includegraphics[width=\textwidth]{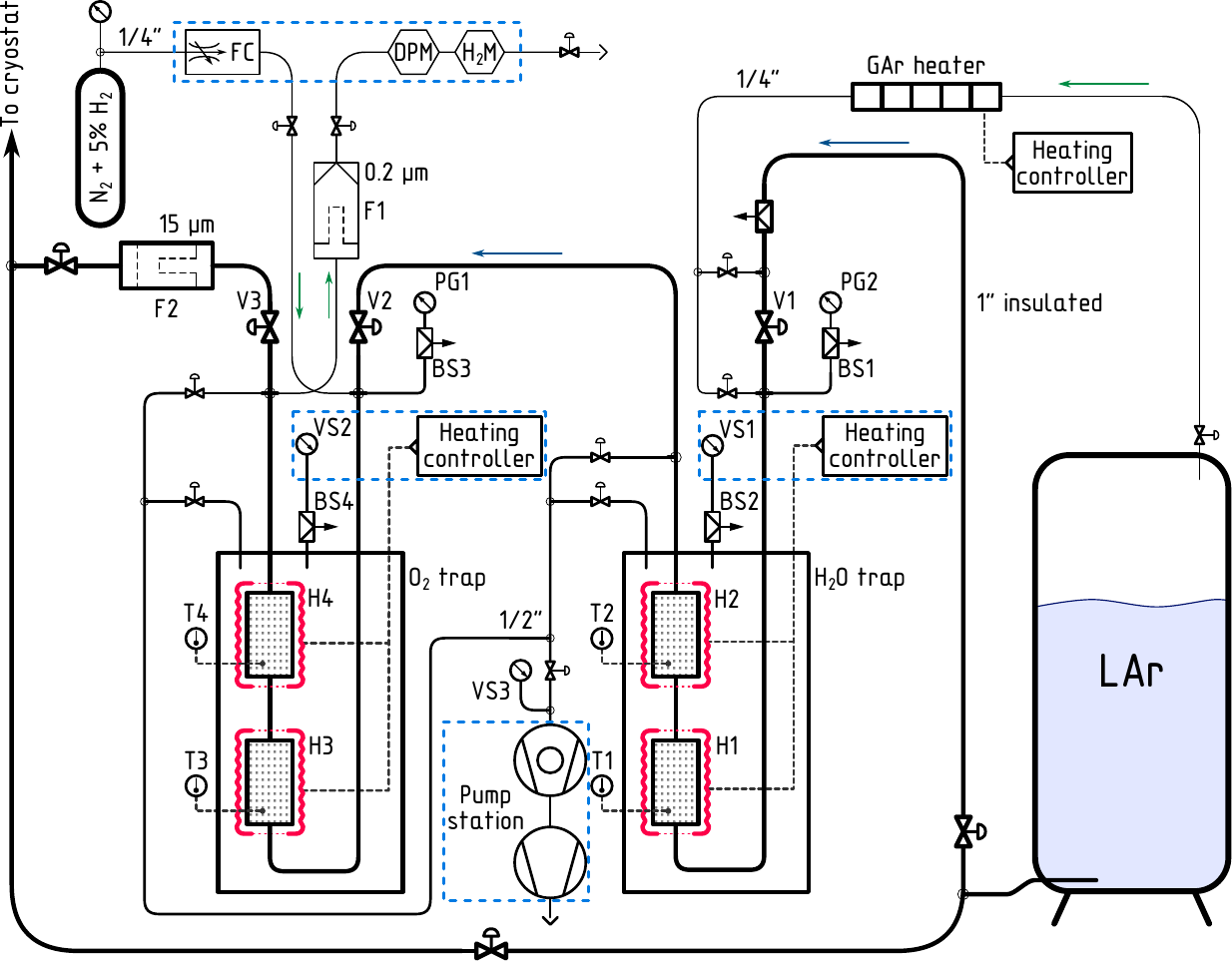}
\includegraphics[width=\textwidth]{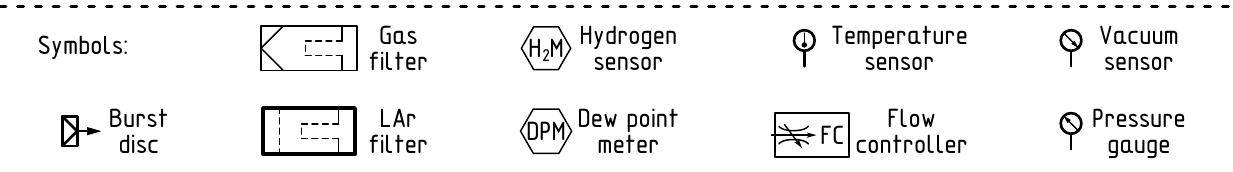}
\caption{\llars~ piping and instrumentation diagram (P\&ID). LAr is pushed from the storage tank on the right through the purification system and into the cryostat on the left (not shown). Both units and their traps are shown, as are the heaters, their controllers, the regeneration gas bottle and its system, and other auxiliary equipment. Please refer to the text for details.}
\label{fig:LLARS_des}     
\end{figure*}

\begin{figure*}[htbp]
\centering
  \includegraphics[width=\linewidth]{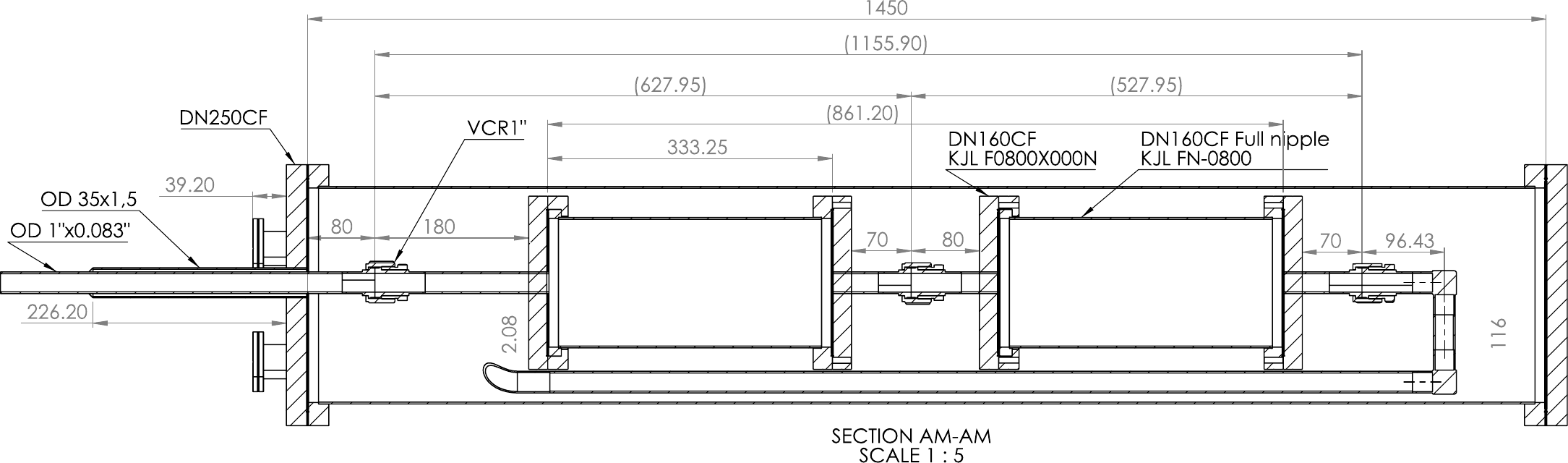}
\caption{Technical drawing of a single unit of the purification system. Two traps based on CF-160 full nipples are housed in a vacuum vessel made from a tube closed with CF-250 flanges. Vacuum provides thermal insulation for the traps. LAr inlet and outlet are located on the CF-250 top flange together with the CF-40 instrumentation and power feedthrough flanges.}
\label{fig:purifier_unit}     
\end{figure*}

The scheme  of the \legend~LAr Purification System (\llars) is shown in Fig.~\ref{fig:LLARS_des}. It consists of two units, each containing two traps connected in series. The first unit has been designed to remove water from LAr, and the second one to remove oxygen. The traps (called H1--H4) are made of CF-160 nipples (\qty{150}{\mm} internal diameter, \qty{333}{\mm} length, \qty{5.9}{\liter} volume) closed on both sides with sintered metal filters with \qty{10}{\micro \m} pore size set in custom-modified CF flanges. The pairs H1--H2 and H3--H4 are installed in vacuum vessels (units) providing thermal insulation. They are based on stainless steel tubes with an internal diameter of \qty{250}{\mm} and are closed on both sides using CF-250 flanges. The inlets (bottoms) of the traps are equipped with a dedicated flow driver (dispenser) that distributes the in-flowing LAr evenly across the filter surface. Sintered metal filters are used to retain the adsorbers in the traps. 

H1--H2 are filled with \qty{4}{\text{\AA}} molecular sieve (\qty{10}{\kg} in total), and H3--H4 with Q-5 copper catalyst, 14 $\times$ 28 mesh beads from Research Catalyst, Inc. This material is equivalent to the “Engelhard Q-5 Reagent” or “BASF Cu-0226”. According to literature, \qty{1}{\kg} of the catalyst can adsorb about {\num{4.7}~{norm.}~L/kg} of gaseous oxygen~\cite{purifiers}. Thus, \qty{1}{\kg} would be sufficient to purify about \qty{5.5}{\cubic \meter} (\qty{7.7}{\tonne}) of LAr, assuming an initial oxygen concentration of \qty{1}{ppm} (L-200 LAr quality). Assuming further that the adsorption properties in the liquid phase are 10 times worse, we use \qty{10}{\kg} of Q-5 catalyst and predict the purification capacity of the system to be around \qty{8}{\tonne}. 

Water adsorption capacity of molecular sieve varies by type, but in general it is about \qty{20}{\percent} of its weight. 
Assuming an efficiency decrease by a factor 10 for liquid phase adsorption again implies that \qty{10}{\kg} of zeolite has a water adsorption capacity of \qty{200}{\g}, which is more than enough to purify \qty{8}{\tonne} of LAr, assuming an initial water concentration of \qty{1}{ppm}. The capacity of the \legend-200 storage tank is \qty{8}{\tonne}, thus the purification system shall be regenerated only after processing at least one full tank. 

The traps H1--H2 and H3--H4 are connected together with \qty{1}{inch} stainless steel tubes. Only metal-sealed joints (VCR, CF) are used to ensure high tightness and robustness against temperature variations.  Each trap is equipped with a dedicated heater (\qty{3.5}{\kilo \W}) and two Pt-100 temperature sensors, one installed in the center of the trap (T1--T4) and the second built into the heater. Each pair of traps shares a burst disc (BS1, BS3) with a burst pressure of \qty{6}{barg}, and an analog pressure gauge (PG1 and PG2). V1, V2, and V3 are cryogenic valves. The vacuum vessels are also equipped with burst discs of the same type (BS2, BS4), and vacuum gauges VS1 and VS2. A technical drawing of one of the units of \llars{} is shown in Fig.~\ref{fig:purifier_unit}.

\subsection{Regeneration procedure}
\label{subsec:regeneration}

The purification agents (copper catalyst and molecular sieve) must be activated prior to their first use, and must be regenerated after saturation or loss of efficiency. In this subsection, we describe the \llars{} regeneration procedures.

The copper catalyst is heated up to \qty{250}{\celsius} core temperature with the electric heaters and flushed with forming gas\footnote{For underground operation at TUM and LNGS, we switched from flammable Arcal-15 (\qty{5}{\percent} hydrogen in argon) to non-flammable forming gas (\qty{5}{\percent} hydrogen in nitrogen).}. There are two controllers that control the heaters via a master-slave system. The slave controller is connected to the Pt-100 sensor installed in the heater, and the master is connected to the temperature sensor located in the center of the trap. In this way, we avoid overheating of both the heater and the adsorbers in the outer layers of the traps.
A sufficiently high temperature induces a reaction, as shown in Eq.~(\ref{eq:cu-regeneration}), creating water that is carried away by the forming gas flow. We set the gas flow between \qty{1}{\liter \per \minute} and \qty{10}{\liter \per \minute} with the flow controller (FC in Fig.~\ref{fig:LLARS_des}) and observe the water content at the exit via the dew point meter (DPM: Superdew 3 system with the Blue dew point sensor). As a safety requirement by LNGS, we also monitored the hydrogen content in the exhaust gas (H$_2$M: Hydrogen sensor BCP-H$_2$ from Bluesens). Once the forming gas flows, the moisture content starts out low, in the single ppm range or below, and increases when the copper reaches about \qty{200}{\celsius}. After roughly \num{900} trap volumes of forming gas are exchanged, the copper oxide is completely reduced to elemental copper, indicated by a decrease in water concentration in the exiting forming gas. As a final step, the trap is pumped to high vacuum and slowly brought back to room temperature.

To regenerate the water trap, we simply heat it to \qty{250}{\celsius} in the same way the copper catalyst is heated, then pump it to high vacuum. Initially, a strong pre-vacuum pump is required to remove the large amount of water released. Later, from around \qty{2e-2}{mbar} (measured by VS3) onward, we switch to a turbo-molecular pump and, once the pressure stops decreasing, slowly return to room temperature. The duration of this process depends heavily on the amount of adsorbed impurities and may take between \qty{24}{\hour} and \qty{36}{\hour}. During normal operation of the system, approximately 8 hours were required to warm up the traps from LAr temperature to \qty{0}{\celsius} when the heaters could be switched on. The controllers do not read negative temperatures correctly and therefore do not work below \qty{0}{\celsius}.

\subsection{Assembly and tests at the Technical University of Munich} 

\begin{figure}[htb]
\centering
  \includegraphics[scale=0.15]{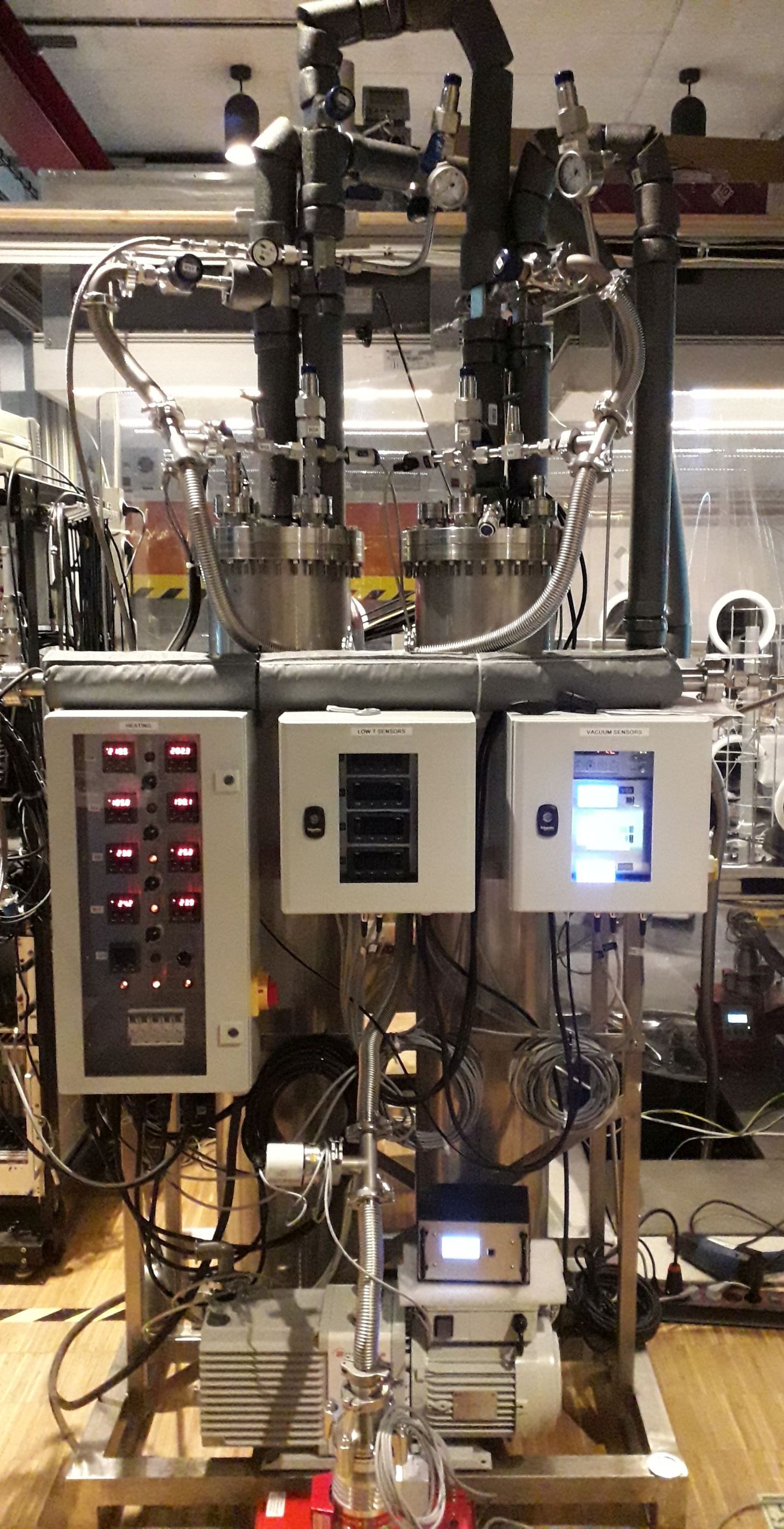}
\caption{\llars\ assembled on a Euro-pallet-sized platform. The large vacuum vessels (purification units) conceal the purification traps within them. The connections to the vessels are located at the top. The foam-insulated LAr lines are easily visible. The electrical control boxes for heating and system monitoring are visible in the front.}
\label{fig:llars}     
\end{figure}

\llars{} has been assembled and initially tested at the Physics Department of the Technical University of Munich (TUM). 
To facilitate transport, the entire purification system was installed on a platform of Euro-pallet size ($\qty{120}{\cm}\times \qty{80}{\cm}$), as shown in Fig.~\ref{fig:llars}. 
At TUM, we operate the Subterranean Cryogenic Argon Research Facility \scarf~\cite{scarf} in a shallow underground laboratory. \scarf\ is a \qty{1}{t} LAr test stand used for R\&D for \legend\ and previously for \textsc{Gerda}. 
We used this facility for the first \llars{} tests, filling \scarf{} with \qty{600}{\kilo \gram} of commercial LAr, and observing the purification performance. 

We received two \qty{600}{\liter} tanks of LAr 5.0 from Air Liquide, with impurity specifications given in Tab.~\ref{tab:air_liquide_TUM_tests}. Argon from the first tank was also sampled with the gas analyzer and the scintillation analyzer. We found impurity concentrations similar to those reported by the vendor. The effective triplet lifetime was measured to be \qty{0.65}{\micro \s}, similar to the lowest previously reported values for LAr 5.0 (see Tab.\ref{tab:triplet_tests}).

\begin{table*}[htbp]
    \centering
    \caption{Certified and measured purities of LAr delivered by Air Liquide to TUM, along with the purity of purified LAr determined by \textsc{Llama} in \scarf.}
    \begin{tabular*}{\textwidth}{@{\extracolsep{\fill}}lccccl@{}} \toprule
               & \multicolumn{3}{c@{}}{Contamination level (ppm)} & & \\ \cmidrule{2-4}
    LAr source & \nit  & \oxy  & \water & $\tau_t$ (\unit{\micro \s}) & Remarks \\ \midrule
    Air Liquide tank 1  & 4.4   & 2.4  &        &               & According to certificate \\
                        & 3.7   & 2.3  &  0.3      & 0.65       & Measured with GA and SA \\ \hline 
    Air Liquide tank 2  & 1.8   & 1.1  &        &               & According to certificate \\ \hline
    \scarf & &  &  & \num{1.30(1)} & Measured with \textsc{Llama} \\ \bottomrule
    \end{tabular*}
    \label{tab:air_liquide_TUM_tests}
\end{table*}

The LAr transport tanks were emptied consecutively, after a complete initial regeneration, through \llars\ into the cryostat without problems. We found that approximately \qty{60}{\kilo \gram} LAr is required to cool the system before the liquid exits its output. Before \llars\ is fully cooled to the operating temperature, mostly flash gas is produced. An overpressure of \qty{1.5}{barg} was sufficient to push LAr through the system, and at a pressure of \qty{3.5}{barg}, the mass flow was around \qty{350}{\kg \per \hour}. This is sufficient for \legend{}-200 as the \qty{8}{\tonne} storage tank would be processed within \qty{22}{\hour}. The insulation vacuum in both units was stable at the level of \qty{2e-2}{mbar}.

To assess the purification performance, we also determined the effective triplet lifetime after the purification process. To do this, an identical twin of \llama\footnote{\textsc{Llama} is introduced in Sec.~\ref{sec:llama}} in \legend{}-200 was set up in \scarf{}. We found an effective triplet lifetime of \qty{1.30(1)}{\micro \s}, exceeding \textsc{Legend-200} specifications\footnote{The \legend-200 specification for LAr is $\tau_t > \qty{1.2}{\micro \s}$} by \qty{0.1}{\micro \s} and demonstrating the system's ability to purify commercial liquid argon efficiently.

\subsection{Integration with LEGEND-200 cryogenic infrastructure}

After completing the assembly and tests at TUM, the purification system was transported to LNGS. It was connected to the \legend-200 LAr storage tank and installed next to it. In this configuration, purification happens in batch mode by using overpressure in the storage tank to push LAr through the purification system and into the cryostat.

At a later time, \llars{} can be moved closer to the cryostat and integrated into a LAr circulation loop. 
A dedicated cryogenic pump was developed and purchased from ILK Dresden (``In\-sti\-tut für Luft- und Käl\-te\-tech\-nik'') and installed at the bottom of the cryostat to pump out LAr. The pump is based on a medium-sized linear drive and features a volume capacity of \qty{700}{\liter \per \hour} at its design frequency of \qty{1}{\hertz}. It has a working frequency range of \qty{0.2}{\hertz} to \qty{1.5}{\hertz} and can produce a pressure difference of up to \qty{2}{bar}.

The pumped-out LAr may be directed via vacuum-insulated tubes to the purification system and returned to the cryostat. 
That way, so-called loop-mode purification can be performed if the LAr quality in the \legend-200 cryostat degrades. 
Loop-mode purification is less effective than batch-mode purification because freshly purified liquid is mixed with the unpurified rest. Assuming perfect mixing and maximal impurity removal, a reduction of $1 / \mathrm{e}$ in impurity concentration is expected for each volume exchange. In realistic conditions, loop-mode purification can take longer. A similar system as \llars{} needed around 20 volume exchanges to elevate the effective triplet lifetime to \qty{1.3}{\micro \s} after an air leak decreased it to \qty{1.0}{\micro \s} in \scarf~\cite{voglpurification}. At its design point, the \legend-200 LAr pump takes \qty{91}{\hour} to circulate the full \qty{64}{\cubic \m} volume of the cryostat. This provides the possibility to perform impactful purification in a moderate time frame, e.g.\ during hardware operations which can last for one or several months.

%% file: 5_L200_filling.tex
\section{Filling of the LEGEND-200 cryostat}
\label{sec:filling}

\subsection{Filling procedure}
 Filling of the \legend-200 cryostat began in June 2021. Initially, \llars\ was fully regenerated following the procedure described in Sec.~\ref{subsec:regeneration}. Subsequently, LAr was pushed from the storage tank through the purification system and guided to the \textsc{Legend-200} cryostat by vacuum-insulated tubes. The storage tank has a capacity of \qty{8.4}{t} (\qty{6}{\cubic \m}) and was refilled twice a week by delivery trucks with L-200 quality LAr. 
 The tank driver applied the clean filling procedure presented in Sec.~\ref{subsec:LAr_qualities_tests_definition} each time, and we supervised the process. During these refills, \llars\ was separated from the storage tank due to significant pressure build-up. Each truck carried a certificate issued by Linde that guaranteed the agreed purity, and each certificate was verified before filling the LAr into \legend.
 
\begin{figure}[htb]
  \includegraphics[width=\linewidth]{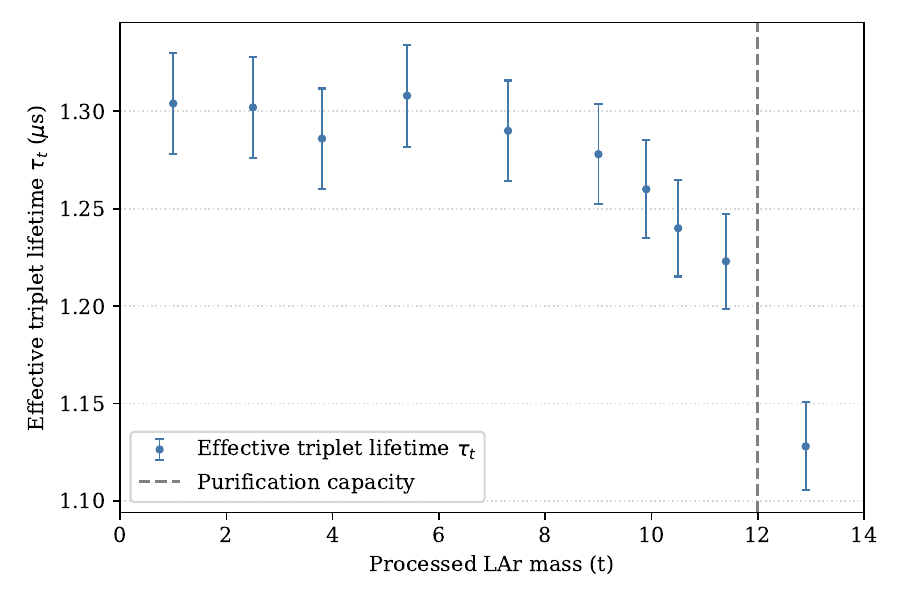}
\caption{The effective triplet lifetime measured with the Scintillation Analyzer is shown versus the cumulative processed L-200 quality LAr mass. A complete regeneration of \llars{} was performed initially. The purification performance decreases after processing \qty{10}{t}, and the \legend-200 effective triplet lifetime threshold of $\tau_t \geq \qty{1.2}{\micro \s}$ is crossed at \qty{12}{t}, which defines the system's capacity. For comparison, measured triplet lifetimes of unpurified L-200 quality LAr range between \qty{0.8}{\micro \s} and \qty{1.0}{\micro\s}.}
\label{fig:saturation}     
\end{figure}

 Initially, less than \qty{100}{\kg \per \hour} were processed to prevent the cryostat from cooling too quickly. 
 Later, the mass flow was increased gradually. The gas analyzer systems permanently monitored the nitrogen, oxygen, and water concentrations in the purified LAr stream. The SA was filled regularly with freshly purified LAr to monitor, in real time, the quality of the purified LAr during filling. Special attention was given before and after the regeneration process of the \llars{} traps, as this step was identified as a potential weak point where contamination could happen. The evolution of the effective triplet lifetime measured by the SA is shown in Fig.~\ref{fig:saturation} as a function of the cumulative processed LAr mass after a complete regeneration, up to the point where the argon quality is no longer acceptable due to the traps saturating. A decrease in the effective triplet lifetime from \qty{1.3}{\micro \s} to about \qty{1.25}{\micro \s} is visible at around \qty{10}{t} of the processed LAr mass. This confirms our theoretical predictions of the needed amount of absorbers (see Sec.~\ref{subsec:design_system}). To reach the LAr purity goal of \textsc{Legend-200} ($\tau_t \geq \qty{1.2}{\micro \s}$), \llars\ can purify slightly more: We determined the purification capacity to be \qty{12}{t} for L-200 quality LAr. 
 Hence, we interrupted the filling to completely regenerate the system always before exceeding a processed mass of \qty{12}{t}, forming an alternating pattern of cryostat filling and system regeneration. An attempt to regenerate only the molecular sieve failed to elevate the effective triplet lifetime above \qty{1.2}{\micro \s}, indicating that the bottleneck is not \ce{H2O} removal. This hypothesis is also supported by the results of the purification tests performed in Krak\'{o}w, when only oxygen removal was applied, see Sec.~\ref{sec:design_adsorbers}.
 
The temperature in the cryostat was monitored with several sensors. Once the cryostat was sufficiently cold, the filling proceeded at up to \qty{400}{\kg \per \hour} (average flow was estimated at \qty{380}{\kg \per \hour}) with the GA continuously operating. Over four weeks, we processed around \qty{100}{t} of LAr, regenerated the system nine times, and filled the cryostat to around \SI{70}{\percent} of its capacity. The filling was then interrupted due to an excessive nitrogen influx (see Sec.~\ref{sec:nitrogen-spoiling}) and resumed at a later point (see Sec.~\ref{sec:top-up}).

\subsection{Monitoring of the LAr quality in the cryostat with \llama}
\label{sec:llama}

The \legend\ Liquid Argon Monitoring Apparatus (\llama{}) is installed at the bottom of the \legend-200 cryostat and continuously monitors the optical quality of the LAr in situ. It became fully operational early during filling, once a sufficient LAr level was reached. Since it measures the LAr at its destination point, it is sensitive to potential sources of impurities anywhere along the line.

\llama\ uses an $^{241}$Am-based triggered scintillation light source in combination with 13 silicon photomultipliers (SiPMs)\footnote{All SiPMs of \llama\ are directly sensitive to the LAr scintillation light wavelength, except for the SiPM in \SI{15}{\centi\meter} distance, which features a fused silica window.} in distances from \SIrange{15}{75}{\centi\meter} from the light source. It simultaneously measures the primary light yield, the effective triplet lifetime of the LAr scintillation, and the attenuation length.

It acquires data in packets of 3~million scintillation events, which takes around 5 hours. The automated processing and analysis, which occur largely in parallel with data taking, yield a set of optical parameters for each packet.
Detailed information about the \llama\ setup, processing, and analysis routines is provided in~\cite{schwarzPhD}.

\begin{figure}[htb]
  \includegraphics[width=\linewidth]{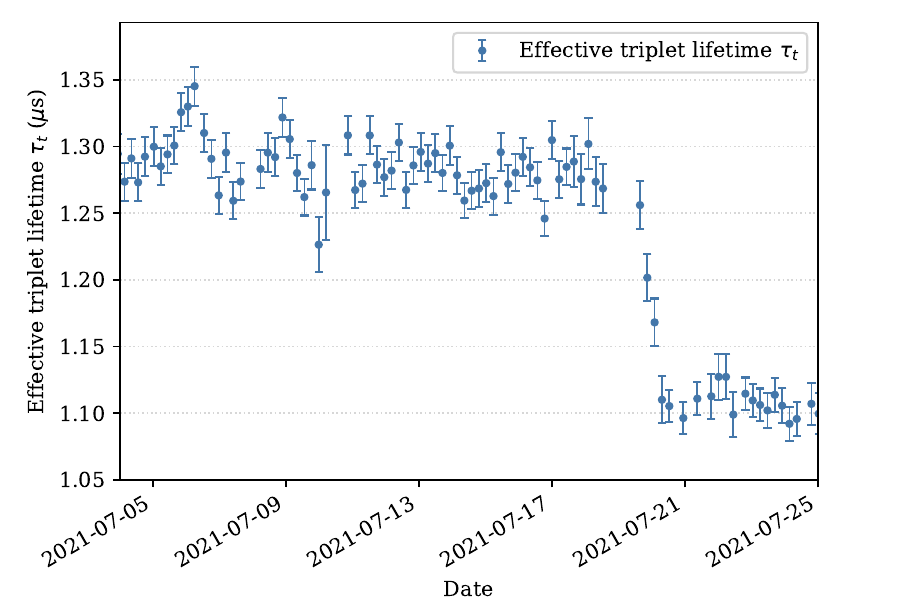}
\caption{Effective triplet lifetime ($\tau_t$) evolution in the cryostat during filling, measured by LLAMA. The sudden drop at the end is caused by nitrogen contamination, discussed in Sec.~\ref{sec:nitrogen-spoiling}.}
\label{fig:llama_data}     
\end{figure}

Fig.~\ref{fig:llama_data} shows the effective triplet lifetime in the cryostat measured by \llama\ during filling. 
The data start slightly above \qty{1.3}{\micro \s}. 
This lifetime is compatible with results obtained during an ultra-high-purity investigation of LAr \cite{heindl2010}, where a hot getter was used to control impurity levels below \qty{10}{ppb}~\cite{heindl2011}.

\subsection{Nitrogen contamination of LAr}
\label{sec:nitrogen-spoiling}

The sudden drop in effective triplet lifetime at the end of the filling campaign (see Fig.~\ref{fig:llama_data}) is due to an influx of low-quality LAr into the \legend-200 cryostat. The last LAr delivery to the storage tank was contaminated with nitrogen, despite being delivered with a proper L-200 quality certificate. The gas analyzer was no longer operational at this point, and the contamination was not detected until after the filling was restarted. The drop in LAr quality was discovered by \llama{}, which led us to immediately\footnote{Filling of contaminated LAr lasted for \SI{17}{\hour} since we demanded several data packets from LLAMA to indicate a degradation before suspending filling. This was done to avoid a false positive caused by data fluctuations.} halt the cryostat filling and investigate the cause of the issue. 

A batch sampled from the storage tank was measured directly on-site with the SA and exhibited a very short effective triplet lifetime of \qty{430}{\nano \s}. Two samples were taken from the storage tank and analyzed in two independent mass spectrometers. The system at TUM yielded a high nitrogen concentration of $[\mathrm{N}_2] = \qty{8.7(34)}{ppm}$ and no signal for oxygen~\cite{voglMSc}. Taking the uncertainties into account, an upper limit of $[\mathrm{O}_2] < \qty{0.97}{ppm}$ can be placed at \qty{90}{\percent}~C.L. The second sample was analyzed at the Jagiellonian University, yielding comparable results ($[\mathrm{N}_2] = \qty{10(2)}{ppm}$). Following this discovery, the vendor certification was reviewed, and a calibration mistake was identified. The calibration was retrospectively corrected, and the contamination with exclusively nitrogen at a level of around \qty{10}{ppm} was confirmed. 
The instrument miscalibration was caused by a power outage, followed by an incorrect initialization.
Additionally, after an outage, the plant usually requires time to reach optimal conditions and cannot immediately provide argon of L-200 grade. Thus, the gas quality was also affected.

To summarize, both mass spectrometers and the vendor measurement indicated contamination with nitrogen at around \qty{10}{ppm}, with no other impurities. This is compatible with the \qty{430}{\nano \s} short effective triplet lifetime measured in the contaminated batch~\cite{warp-nitrogen}\footnote{To reach the low effective triplet lifetime measured in the unpurified LAr, one has to assume the presence of additional impurities. For example, \SI{0.85}{ppm} oxygen would be within limits and would lead to the measured expected triplet lifetime in combination with the observed nitrogen concentration.}.

This nitrogen incident demonstrates that in-flow gas analyzers are crucial for quality monitoring and that a device like \llama{} is essential to ensure high LAr quality, especially during filling. 
\llama{}'s capabilities as an early warning system allowed the contamination to be controlled, leading to a negligible degradation of the LAr instrumentation's physics performance~\cite{schwarzPhD}.

\subsection{Final top up}
\label{sec:top-up}

\begin{figure}[htb]
    \centering
    \includegraphics[width=\linewidth]{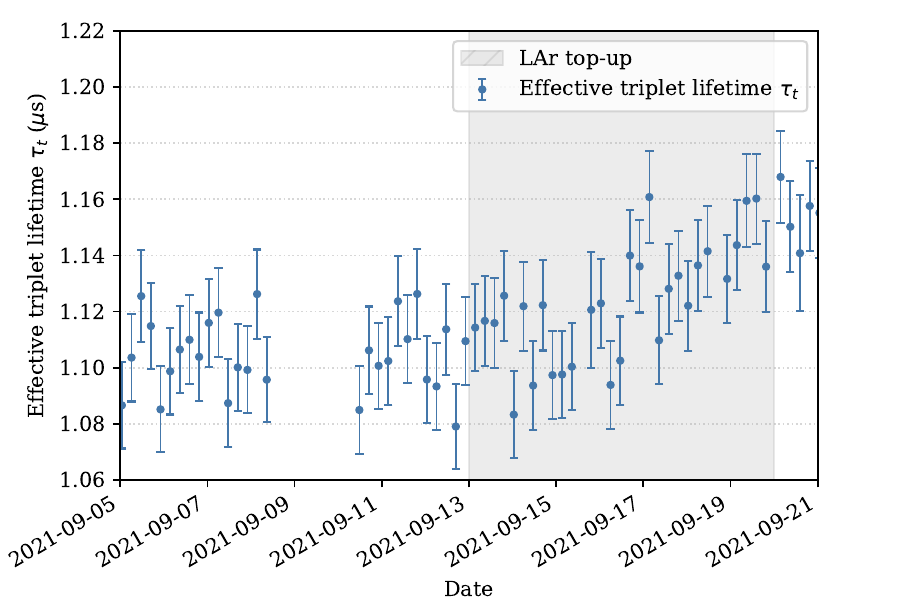}
    \caption{Effective triplet lifetime ($\tau_t$) evolution before and during the final LAr top-up measured with \llama{}. The LAr was purified during filling and contained fewer impurities than the liquid already present in \legend-200. Thus, the impurities were diluted, and the scintillation properties improved.}
    \label{fig:triplet-evo-topup}
\end{figure}

The contaminated batch led to a premature halt of the \textsc{Legend-200} filling process, with approximately \qty{60}{t} filled out of \qty{91}{t}. The situation was resolved after a two-month break. The vendor removed the contaminated LAr from the storage tank and delivered a new batch with proper purity and a certificate. We topped up the cryostat with purified argon of excellent quality. As a result, the impurity concentration slightly decreased, and the optical properties improved. An evolution plot of the effective triplet lifetime before and during the top-up is shown in Fig.~\ref{fig:triplet-evo-topup}.
The top-up concluded the \legend-200 LAr filling campaign with a final effective triplet lifetime in the cryostat of around \SI{1.16}{\micro \second}~\cite{schwarzPhD}.

\subsection{Nitrogen-doped argon scintillation analysis}

Data collected with \llama\ around and during the nitrogen spoiling event provide insights into the effects of nitrogen impurities on LAr scintillation. Although dedicated measurements have already been conducted~\cite{warp-nitrogen}, the data at hand prove valuable due to the sub-ppm concentrations reached. For details of the analysis and the results, the reader is referred to \cite{schwarzPhD}. Additionally, a dedicated report is in preparation.

\begin{figure}[h]
  \includegraphics[width=\linewidth]{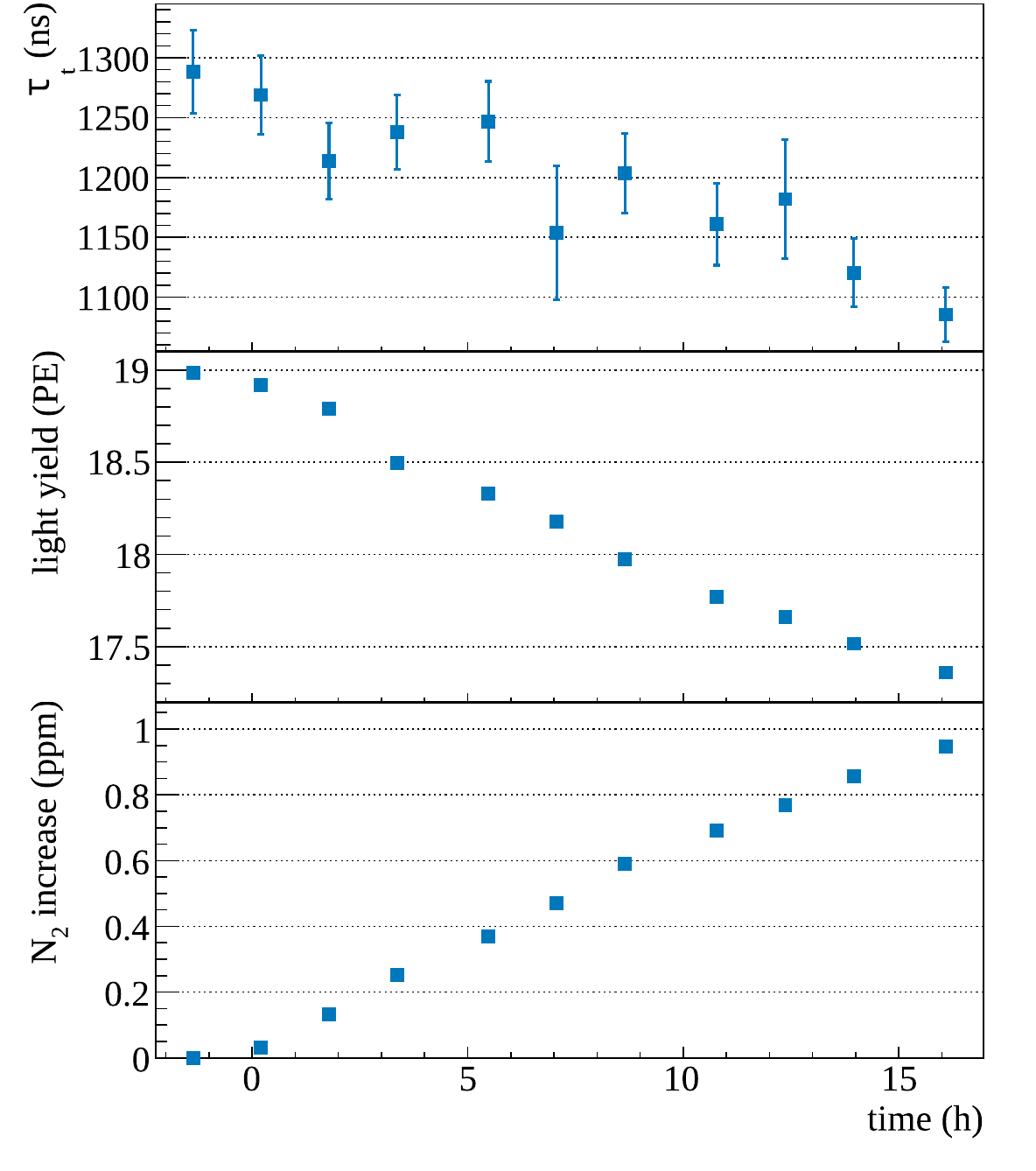}
\caption{Time resolved evolution of the effective triplet lifetime of LAr scintillation (top) and the primary light yield (center) together with the calculated increase in nitrogen concentration (bottom). 
Overall, the nitrogen content increased by \SI{0.9}{ppm} over the course of \SI{17}{hours}. 
Taken from \cite{schwarzPhD}.}
\label{fig:doping_timeslices}     
\end{figure}

The data obtained during the filling of the contaminated LAr provide a correlation between the effective triplet lifetime of the LAr scintillation, the primary light yield, and the nitrogen concentration. The time-resolved evolution of these parameters is shown in Fig.~\ref{fig:doping_timeslices}.

While the effective triplet lifetime and the primary light yield are obtained by LLAMA, the increase in nitrogen concentration is calculated. The calculation uses parameters of the filling process, such as the LAr transfer rate and the LAr content of the cryostat immediately after the filling was halted. The derivation uses the assumption that LLArS' nitrogen removal capabilities were negligible during the filling of the spoiled LAr batch, as we expect it to be almost saturated with nitrogen before the filling of said batch commenced.
We derive the nitrogen concentration in the \legend-200 cryostat after filling the spoiled LAr to be \qty{0.9}{ppm}.

\subsection{Nitrogen removal capability of LLArS}

In a dedicated test, we investigated the capability of LLArS to remove traces of nitrogen diluted in LAr. 
To this end, we processed around \SI{100}{\liter} of nitrogen-contaminated LAr from the storage tank with LLArS, immediately after a full regeneration cycle. The resulting LAr was measured with the scintillation analyzer, yielding an effective triplet lifetime of $\tau_t \approx \SI{800}{\nano\second}$. In contrast, a measurement of the unpurified LAr ($\approx \SI{10}{ppm}$ N$_2$) resulted in $\tau_t \approx \SI{430}{\nano\second}$. This suggests a limited capability of LLArS to trap nitrogen contaminants if its concentration is high ($\sim$ 10 ppm).

%% file: 6_conclusions.tex
\section{Conclusions}
\label{sec:conclusions}

\legend-200 requires LAr with sufficient purity to achieve its physics goals. Commercial LAr with 6.0 purity might not meet the requirements, as concentrations of \nit{}, \oxy{}, and \water{} can reach \SI{0.5}{ppm}. It was therefore decided to perform liquid-phase purification of delivered LAr during the filling of the \legend-200 cryostat.

At the vendor's production plant, we verified the quality of the produced LAr and established a delivery chain to LNGS, which was subsequently tested successfully. Together with the vendor, we defined a custom ``L-200 quality'' LAr specification with oxygen and water concentrations below \SI{1.0}{ppm}, and nitrogen below \SI{1.4}{ppm} to optimize the performance of the purification system.

The custom LLArS purification system was tested at TUM and successfully employed at the \legend-200 filling campaign.
LLArS achieves a high LAr purity, with effective triplet lifetimes reaching up to \SI{1.3}{\micro\second}. After the initial cool-down phase, an average processing speed of \SI{380}{\kilo\gram\per\hour} was reached. Three independent quality control systems assessed the LAr purity during the filling.

The filling was interrupted due to indications of degrading LAr quality from LLAMA, caused by a nitrogen-contaminated LAr delivery. Due to the nitrogen impurity, the final effective triplet lifetime is \SI{1.16}{\micro\second} after reaching the target LAr content of \SI{91}{\tonne}.

LLArS is stationed at LNGS to enable infrequent top-ups of the cryostat content with purified LAr whenever the fill level drops significantly. If the LAr quality deteriorates drastically, LLArS can be incorporated in a closed loop to re-purify the LAr in the cryostat. We plan to adopt LLArS to purify LAr for the \legend{}-1000 cryostat after the \legend{}-200 stage is finished.